\pdfoutput=1
\documentclass[longauth]{aa} 
\usepackage[pdftex]{graphicx}
\usepackage[pdftex]{graphics}
\usepackage{colortbl}
\usepackage{url}
\usepackage{txfonts}
\usepackage{natbib}
\bibpunct{(}{)}{;}{a}{}{,} 
\usepackage{amssymb}
\usepackage{subfigure}

\begin{document}
\renewcommand{\topfraction}{0.85}
\renewcommand{\bottomfraction}{0.7}
\renewcommand{\textfraction}{0.15}
\renewcommand{\floatpagefraction}{0.66}

\newcommand{\HMS}[3]{$#1^{\mathrm{h}}#2^{\mathrm{m}}#3^{\mathrm{s}}$}
\newcommand{\DMS}[3]{$#1^\circ #2' #3''$}
\newcommand{\degree}{$^\circ$}
\newcommand{\zetahard}{$\zeta_{\mathrm{hard}}$}
\newcommand{\zetastd}{$\zeta_{\mathrm{std}}$}
\newcommand{\UNITS}[1]{\,\mathrm{#1}}
\newcommand{\hess}{H.E.S.S.}
\newcommand{\hessstdana}{H.E.S.S. Standard Analysis}
\newcommand{\HESSJ}{HESS~J1646--458}
\newcommand{\FRMSRC}{0FGL~J1648.1--4606}
\newcommand{\PSRJ}{PSR\,J1648--4611}
\newcommand{\GX}{GX340+0}
\newcommand{\g}{$\gamma$}
\newcommand{\order}{$\mathcal{O}$}
\newcommand{\TODO}[1]{\textbf{\texttt{\textcolor{green}{\emph{[#1]}}}}}

\newcommand{\MS}{$M_\odot$}
\newcommand{\TODOMF}[1]{\textbf{\texttt{\textcolor{red}{\emph{[#1]}}}}}

\title{Discovery of extended VHE \g-ray emission from the vicinity 
of the young massive stellar cluster Westerlund~1}

\author{HESS Collaboration
\and A.~Abramowski \inst{1}
\and F.~Acero \inst{2}
\and F.~Aharonian \inst{3,4,5}
\and A.G.~Akhperjanian \inst{6,5}
\and G.~Anton \inst{7}
\and A.~Balzer \inst{7}
\and A.~Barnacka \inst{8,9}
\and U.~Barres~de~Almeida \inst{10}\thanks{supported by CAPES Foundation, Ministry of Education of Brazil}
\and Y.~Becherini \inst{11,12}
\and J.~Becker \inst{13}
% ...
}

\offprints{Stefan Ohm, \email{physoh@leeds.ac.uk}, 
  Milton Virg{\'{\i}}lio Fernandes, \email{milton.virgilio.fernandes@physik.uni-hamburg.de}}

\institute{
Universit\"at Hamburg, Institut f\"ur Experimentalphysik, Luruper Chaussee 149, D 22761 Hamburg, Germany \and
Laboratoire Univers et Particules de Montpellier, Universit\'e Montpellier 2, CNRS/IN2P3,  CC 72, Place Eug\`ene Bataillon, F-34095 Montpellier Cedex 5, France \and
Max-Planck-Institut f\"ur Kernphysik, P.O. Box 103980, D 69029 Heidelberg, Germany \and
Dublin Institute for Advanced Studies, 31 Fitzwilliam Place, Dublin 2, Ireland \and
National Academy of Sciences of the Republic of Armenia, Yerevan  \and
Yerevan Physics Institute, 2 Alikhanian Brothers St., 375036 Yerevan, Armenia \and
Universit\"at Erlangen-N\"urnberg, Physikalisches Institut, Erwin-Rommel-Str. 1, D 91058 Erlangen, Germany \and
Nicolaus Copernicus Astronomical Center, ul. Bartycka 18, 00-716 Warsaw, Poland \and
CEA Saclay, DSM/IRFU, F-91191 Gif-Sur-Yvette Cedex, France \and
University of Durham, Department of Physics, South Road, Durham DH1 3LE, U.K. \and
Astroparticule et Cosmologie (APC), CNRS, Universit\'{e} Paris 7 Denis Diderot, 10, rue Alice Domon et L\'{e}onie Duquet, F-75205 Paris Cedex 13, France \thanks{(UMR 7164: CNRS, Universit\'e Paris VII, CEA, Observatoire de Paris)} \and
Laboratoire Leprince-Ringuet, Ecole Polytechnique, CNRS/IN2P3, F-91128 Palaiseau, France \and
Institut f\"ur Theoretische Physik, Lehrstuhl IV: Weltraum und Astrophysik, Ruhr-Universit\"at Bochum, D 44780 Bochum, Germany
% ...
}

\date{Received 23 August 2011 / Accepted 12 October 2011}

\abstract
{} 
{ Results obtained in very-high-energy (VHE; $E\ge100~$GeV)
  $\gamma$-ray observations performed with the \hess\ telescope array
  are used to investigate particle acceleration processes in the
  vicinity of the young massive stellar cluster Westerlund~1 (Wd~1).
}
{ Imaging of Cherenkov light from $\gamma$-ray induced particle
  cascades in the Earth's atmosphere is used to search for VHE
  $\gamma$ rays from the region around Wd\,1. Possible catalogued
  counterparts are searched for and discussed in terms of morphology
  and energetics of the \hess\ source.  }
{ The detection of the degree-scale extended VHE $\gamma$-ray source
  \HESSJ\ is reported based on 45\,hours of H.E.S.S. observations
  performed between 2004 and 2008. The VHE \g-ray source is centred on
  the nominal position of Wd~1 and detected with a total statistical
  significance of $\sim20\,\sigma$. The emission region clearly
  extends beyond the H.E.S.S. point-spread function (PSF). The
  differential energy spectrum follows a power law in energy with an
  index of $\Gamma=2.19 \pm 0.08_{\mathrm{stat}} \pm
  0.20_{\mathrm{sys}}$ and a flux normalisation at 1\,TeV of $\Phi_0$
  = $(9.0 \pm 1.4_{\mathrm{stat}} \pm
  1.8_{\mathrm{sys}})~\times~10^{-12}
  \UNITS{TeV^{-1}\,cm^{-2}\,s^{-1}}$.  The integral flux above
  0.2\,TeV amounts to $(5.2 \pm 0.9)~\times~10^{-11}
  \UNITS{cm^{-2}\,s^{-1}}$.}
{ Four objects coincident with \HESSJ\ are discussed in the search of
  a counterpart, namely the magnetar CXOU~J164710.2--455216, the X-ray
  binary 4U~1642--45, the pulsar PSR~J1648--4611 and the massive
  stellar cluster Wd~1. In a single-source scenario, Wd~1 is favoured
  as site of VHE particle acceleration. Here, a hadronic parent
  population would be accelerated within the stellar cluster. Beside
  this, there is evidence for a multi-source origin, where a scenario
  involving PSR~J1648--4611 could be viable to explain parts of the
  VHE \g-ray emission of \HESSJ.}

\authorrunning{\hess\ Collaboration}
\titlerunning{VHE \g-ray emission from the vicinity of Westerlund~1}
\keywords{Galaxy: open clusters and associations -- Galaxy: individual objects : 
  \object{Westerlund~1} -- gamma rays: observations -- galaxies: star clusters }

\maketitle

%
%

\section{Introduction}\label{section:introduction}
The long-standing question on the origin and acceleration mechanisms
of hadronic and leptonic Galactic cosmic rays (GCRs) is still not
settled, despite considerable progress. The detection of
very-high-energy (VHE) \g-ray emission from shell-type supernova
remnants (SNRs), e.g. Cassiopeia A, RX\,J1713--3946, RX~J0852.0--4622,
RCW~86, SN~1006 \citep[summarised in][]{Hinton2009+}, and recently
HESS~J1731--347 \citep{HESS:1731} and Tycho's SNR \citep{Acciari2011+}
supports the widely accepted idea of SNRs being acceleration sites of
GCRs. It has been noted for many years that the Galactic SNR
population provides sufficient energy input to sustain the CR flux
measured at Earth. The underlying theory assumes that electrons and
protons are injected into SNR shock fronts where they are accelerated
via the diffusive shock acceleration process up to energies of
$\sim$10$^{15}$\,eV
\citep{Shocks:Krymskii,Shocks:Axford,Shocks:Bell,Shocks:Blandford}.
The ability of SNRs to accelerate electrons up to the so-called knee
in the differential energy spectrum of the GCRs and our common belief
that this holds for protons, too, constitute the paradigm that SNRs
are the long-thought sources of GCRs. In interactions with the ambient
medium, i.e. matter and electromagnetic fields, these GCRs then
produce VHE \g\ rays which can be detected by current imaging
atmospheric Cherenkov telescope (IACT) systems, e.g. \hess, MAGIC,
VERITAS or CANGAROO-III. Additionally, evolving SNRs could explain the
chemical composition up to the knee region. Furthermore, core-collapse
supernovae could explain observed overabundances of some isotopes,
e.g. $^{22}$Ne \citep{Higdon2003+}. However, recent studies applied to
RX\,J1713--3946 highlight potential problems for a dominant hadronic
interpretation for this object \citep{Ellison2010+} and motivate the
search for other acceleration sites and processes.

SNR shells are not the only sites in the Galaxy where GCRs can be
produced via diffusive shock acceleration. One alternative scenario is
particle acceleration in strong shocks in colliding wind binaries
(CWBs). Massive stars are to a large extent bound in binary systems
\citep[e.g.][]{Zinnecker03,Gies08}, generally exhibit high mass-loss
rates ($10^{-5}$\,M$_{\odot}$\,yr$^{-1}$ --
$10^{-3}$\,M$_{\odot}$\,yr$^{-1}$) and drive strong supersonic winds
with velocities of the order of a few $10^3$\,km\,s$^{-1}$.  When
these winds collide in a stellar binary system they form a wind-wind
interaction zone where charged particles can be accelerated to high
energies \citep[e.g.][]{Eichler93}. Electrons can then up-scatter
stellar photons present in the wind collision zones via the inverse
Compton (IC) process to GeV energies
\citep{Muecke02,Manolakou2007+}. On the other
hand, relativistic nucleons can inelastically scatter with particles
in the dense wind and produce $\pi^{0}$s which subsequently decay into
VHE \g\ rays \citep{Benaglia03,CWB:Bednarek05,Domingo2006+,Reimer06}. Apart from
acceleration in binaries, GCRs can be accelerated in the winds of
single massive stars \citep[e.g.][]{Montmerle1979}.

Another scenario involves collective stellar winds: It is commonly
accepted that the bulk (if not all) of the core-collapse SN progenitor
stars and CWBs evolve from collapsing gas condensations in giant
molecular clouds \citep[e.g.][]{Zinnecker07} and mostly remain close
to their birthplaces in groups of loosely bound associations or dense
stellar clusters. When the winds of multiple massive stars in such
systems collide they form a collective cluster wind which drives a
giant bubble (\order(100\,pc)), also referred to as \emph{superbubble}
(SB), filled with a hot ($T \approx 10^6$\,K) and tenuous ($n <
1$\,cm$^{-3}$) plasma \citep[e.g.][]{Weaver77,Silich2005+}. At the
wind interaction zones, e.g. at the termination shock of the stellar
cluster wind, turbulences in form of magneto-hydrodynamic (MHD)
fluctuations and weak reflected shocks can build up.  Unlike SNR shock
fronts and CWBs where GCRs are accelerated through the 1$^\mathrm{st}$
order Fermi acceleration, turbulences in SB interiors can accelerate
particles to very high energies also via the 2$^\mathrm{nd}$ order
Fermi mechanism \citep[e.g.][]{Bykov01}. Moreover, after a few million
years, supernova explosions of massive stars ($M>8$\,M$_\odot$) in the
thin and hot SB environment eventually lead to efficient particle
acceleration at the boundary of the SB or at MHD turbulences and
further amplify existing MHD turbulences \citep[e.g.][and references
therein]{Ferrand2009+}. The interaction of these GCRs with the ambient
medium including molecular clouds or electromagnetic fields leads to
the production of VHE \g\ rays which can then be studied on Earth.
Therefore, stellar clusters are promising targets to study
acceleration and propagation processes of GCRs.

One of the most prominent objects among stellar clusters in the Galaxy
is Westerlund 1 (Wd 1). After its discovery in 1961
\citep{Westerlund61} subsequent observations have established Wd~1 as
the most massive stellar cluster in our Galaxy placing a lower limit
on its mass of $10^5$\,M$_\odot$ \citep{Clark2010+}. An unprecedented
accumulation of evolved massive stars is found without indication of
the presence of an early-type main-sequence star. Amongst the most
massive stars, 24 Wolf-Rayet stars (binary fraction $\geq62$\%) have
been detected and a number of $\sim$150 OB super- and hypergiants
(binary fraction $\sim30$\%) is expected \citep[][and references
therein]{Crow06,Dougherty2010+}.

The analysis of \emph{Chandra} data revealed an arc minute-scale
extended diffuse X-ray emission \citep{Muno06b} which is only seen for
a few young stellar associations in the Galaxy, for example RCW\,38
\citep{Wolk02} and possibly the Arches cluster \citep{Law04} as well
as in the Large Magellanic Cloud in 30 Doradus C \citep{Bamba04} and
DEM L192 \citep{Cooper04}. The total X-ray luminosity of the observed
diffuse emission within $5'$ of Wd\,1 is dominated by its non-thermal
component and amounts to $L_\mathrm{X} \approx 3 \times
10^{34}$\,erg\,s$^{-1}$ which represents just a fraction of 10$^{-5}$
of the total mechanical power in this system \citep{Muno06b}.
However, models as in \citet{Oskinova2005} predict a thermal X-ray
luminosity of $\sim 10^{37}$\,erg\,s$^{-1}$ for stellar clusters
comparable to Wd\,1, which was clearly not observed by \emph{Chandra}
for Wd~1. As for previous observations, there remains the open
question into which channel most of the unobserved energy is
dissipated.

The detection of VHE \g-ray emission from \HESSJ\ was initially
reported in \citet{Wd1:Ohm10} and \citet{HESS:Wd1}. This paper focuses
on a detailed spectral and morphological study of the emission region
and investigates a possible multi-source origin. An in-depth search
for plausible counterparts is conducted and possible
acceleration-mechanism scenarios are elaborated.

%
%

\section{\hess\ Observations and Data Analysis}

Given the large $\sim5^\circ$ field of view (FoV) combined with the good off-axis 
sensitivity, observations with \hess\ are perfectly suited to cover the vicinity of 
Wd~1 and allow for the detailed morphological study of extended sources such as 
\HESSJ. Thereby any large-scale non-thermal VHE \g-ray emission around Wd~1 can be 
probed.

\subsection{The \hess\ Experiment}
The High Energy Stereoscopic System (\hess) is an array of four imaging 
atmospheric Cherenkov telescopes located in the Khomas Highland of Namibia, 1800\,m 
above sea level. The telescopes are identical in construction and each one is 
comprised of a 107\,m$^2$ optical reflector composed of segmented spherical mirrors. 
These focus the incident light into a fine-grained camera built of 960 
photomultiplier tubes. By means of the \emph{imaging atmospheric Cherenkov 
technique} \citep[see e.g.][]{Hillas85} Cherenkov light, emitted by the 
highly-relativistic charged particles in extensive air showers, was imaged by 
the mirrors onto the Cherenkov camera. A single shower was recorded by multiple 
telescopes under different viewing angles. This allowed for the stereoscopic 
reconstruction of the primaries' direction and energy with an average energy 
resolution of 15\% and an event-by-event angular resolution better than 0.1\degree\ 
\citep[Gaussian standard deviation,][]{HESS:Crab}. 

\subsection{The Data Set}

The region around Wd~1 was observed during the H.E.S.S. Galactic Plane
Survey (GPS) in 2004 and 2007
\citep{HESS:SP,HESS:SPChaves}. Additionally, follow-up observations
pointing in the direction of Wd~1 have been performed from May to
August 2008. Data taken under unstable weather conditions or with
malfunctioning hardware were excluded in the standard data quality
selection procedure \citep{HESS:Crab}. Also, pointed observations on
Wd~1 at very large zenith angles of more than $55$\degree\ were
excluded due to systematic effects in the description of the camera
acceptance at such low altitudes for an extended source like
\HESSJ. After quality selection and dead time correction the total
observation time of 45.1\,hours was reduced to a live time of
33.8\,hours. Observations have been carried out at zenith angles from
21\degree\ to 45\degree\ with a mean value of 26\degree\ and an
average pointing offset from the Wd~1 position of 1.1\degree.

\subsection{Analysis Technique}

The data set presented here was processed using the \hessstdana\ for
shower reconstruction \citep{HESS:Crab} and the Boosted Decision Trees
(BDT) method to suppress the hadronic background component
\citep{TMVA}\footnote{The software which was used to analyse the VHE
  \g-ray data presented in this work is the \emph{H.E.S.S. analysis
    package (HAP)} in version 10-06-pl07.}. By parametrising the
centre of gravity and second moments of the recorded extensive air
shower image \citep{Hillas85} in multiple telescopes the shower
geometry of the incident primary particle was reconstructed
stereoscopically. The directional information together with the
measured image intensity was used to reconstruct the energy of the
event.
Since observations have been conducted over four years, the optical
reflectivity of the \hess\ mirrors varied and the gains of the
photomultipliers changed. This effect has been taken into account in
the spectral reconstruction by calibrating the energy of each event
with single muon rings \citep{HESS:Crab}. The decision tree-based
machine learning algorithm BDT returns a continuous variable called
$\zeta$ which was used to select \g-ray-like events. Cutting on this
parameter results in an improvement in terms of sensitivity compared
to the \hessstdana\ of $\sim 20$\% and $\sim 10$\% for spectral
and morphological analysis, respectively \citep{TMVA}.

Similar to the \hessstdana, two sets of \g-ray selection cuts have
been defined in \citet{TMVA}. For the production of sky images the
\emph{\zetahard-cuts} are used.  They require a minimum intensity of
160 photo electrons (p.e.) in each camera image yielding a superior
angular resolution of less than 0.1\degree\ even at large offsets of
2.5\degree\ from the telescope pointing position. Additionally, 30\%
more background events are rejected resulting in a 10\% higher
sensitivity compared to the \hessstdana. For the spectral analysis a
low energy threshold is desirable for a broad energy coverage and
achieved by applying the \emph{\zetastd-cuts} with a 60~p.e. cut on
the image intensity. For the data set under study, this infers an
energy threshold of 450\,GeV for spectral analysis and 700\,GeV for
morphological analysis.

For two-dimensional sky image generation and morphology studies, the
\emph{template} background model \citep{Rowell03,HESS:Background} is
applied. For this method the CR background is estimated in parameter
space rather than in angular space. In this analysis, the BDT output
parameter $\zeta$ has been used to define signal and background
regions. The normalisation $\alpha$ between signal and background is
calculated as the fraction of all events in the FoV falling into the
signal regime, excluding source regions, divided by the number of all
events in the FoV in the background regime, again excluding all source
regions. The system acceptance to measure \g-ray like and CR-like
events drops off radially with the distance to the telescope pointing
position. Since this acceptance is different for both types of events,
a correction is applied to $\alpha$.  Sky images obtained with the
\emph{template} background model agree with sky images generated with
the \emph{ring} background method. The \emph{ring} background method
estimates the signal-like hadronic CR contribution at each trial
position on the sky by integrating events in an annulus centred on
that position, excluding potential source regions.

\begin{figure}[t]
  \centering
  \resizebox{\hsize}{!}{\includegraphics{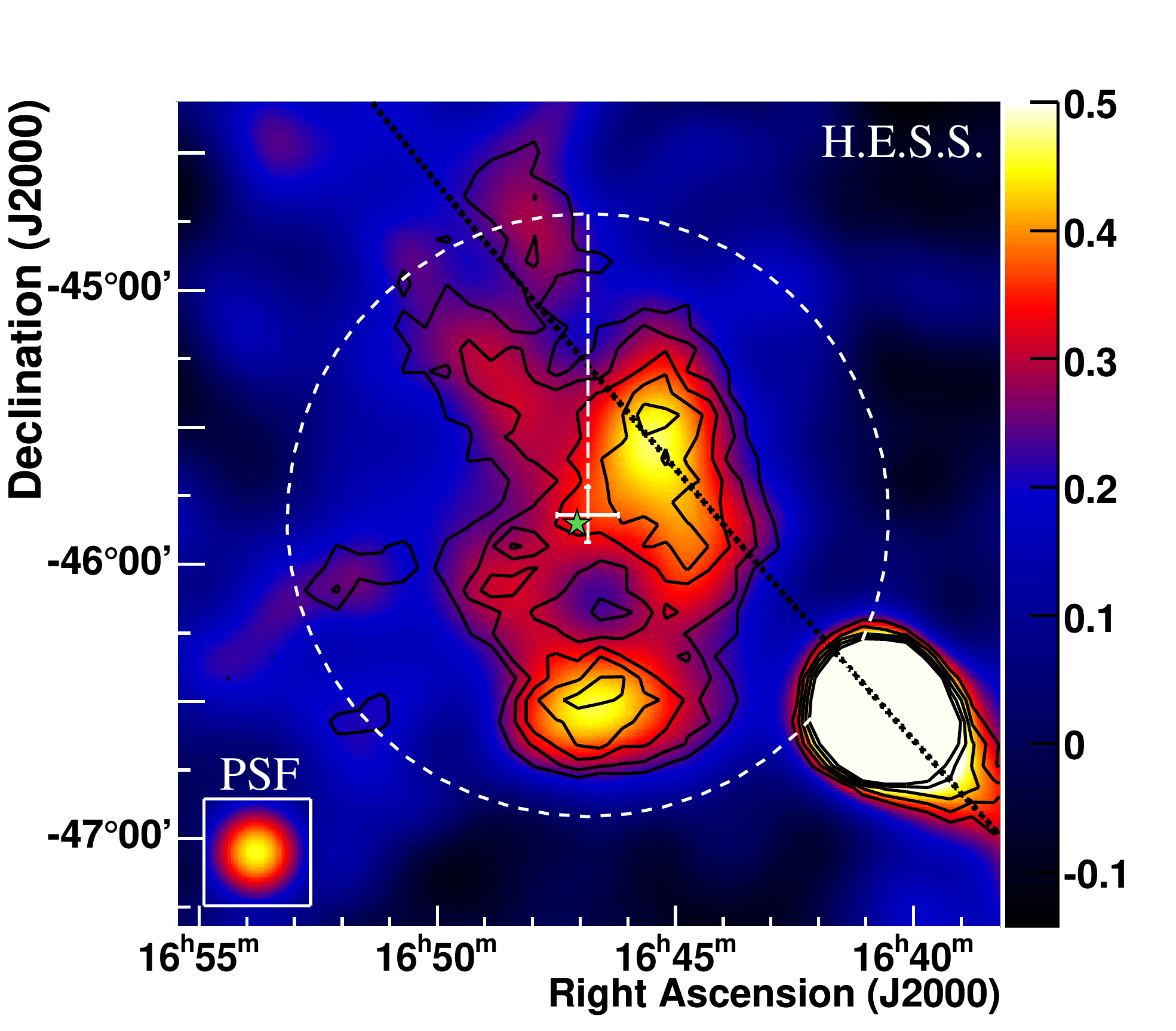}}
  \caption{\hess\ excess map of the region around Wd~1 corrected for
    the camera acceptance, in units of equivalent on-axis VHE \g-ray
    events per arcmin$^2$ and obtained with the \emph{template}
    background method. The image is smoothed with a 2D Gaussian kernel
    with a variance of 0.12\degree\ to reduce the effect of
    statistical fluctuations.  Significance contours between
    4\,$\sigma$ and 8\,$\sigma$ are overlaid in black, obtained by
    integrating events within a radius of 0.22 degrees at each given
    position. The green star marks the position of Wd~1, the white
    cross the best fit position of the VHE $\gamma$-ray emission and
    the black dashed line the Galactic plane. The inlay in the lower
    left corner represents the size of a point-like source as it would
    have been seen by \hess\ for this analysis and the same smoothing,
    normalised to the maximum of \HESSJ. The dotted white circle has a
    radius of 1.1$^\circ$ and denotes the region which was used for
    the spectral reconstruction of the VHE \g-ray emission. Note that
    the bright region in the lower right corner is the source
    HESS~J1640--465 detected during the GPS \citep{HESS:SP}.  }
\label{fig:VHE}
\end{figure}

Figure~\ref{fig:VHE} shows the VHE $\gamma$-ray count map of the
region around Wd~1 and reveals very extended \g-ray emission. The
complex morphology apparent in the sky image and a potential
multi-source origin of the emission is investigated and discussed in
detail in Section\,\ref{sec:results}. Since it is not possible to
estimate the background from the same FoV due to the fact that
observations have been carried out within regions of VHE \g-ray
emission, the \emph{On-Off} background estimation method is utilised
to extract spectral information for the whole emission region,
indicated as a white circle in Figure~\ref{fig:VHE}. Here, the CR
background is subtracted from the source region (\emph{On data}) using
extragalactic observations taken without any VHE \g-ray signal in the
FoV (\emph{Off data}). To ensure similar observational conditions for
\emph{On} and \emph{Off data}, only \emph{On-Off} pairs of
observations are considered that were taken at similar zenith angles
and within four months of each other, resulting in a total live time
of 20~hours for the \emph{On data} set used for spectral analysis. The
absolute normalisation $\alpha$ between \emph{On} and \emph{Off data}
is calculated using the fraction of total events in both observations
(again, excluding potential source regions). The \emph{reflected}
background method \citep{HESS:Background} is used to derive spectral
information for smaller regions and the full data set as discussed in
Section~\ref{subsection:morphology} and \ref{subsection:spectrum}.

All studies presented in this work were cross-checked by a second
analysis chain which is based on the H.E.S.S. standard event
reconstruction scheme \citep{HESS:Crab} using the Hillas second moment
method (Hillas 1985) and an independent calibration of pixel
amplitudes and identification of problematic or dead pixels in the
IACT cameras. Additionally, the \emph{Model Analysis} \citep{Model++}
for the selection of \g-ray-like events has been utilised to
cross-check the spectral results. All analyses give compatible
results.

\section{VHE Results}\label{sec:results}

\subsection{Position}\label{subsection::position+spectrum}

\begin{figure}[t]
  \centering
      \resizebox{\hsize}{!}{\includegraphics[]{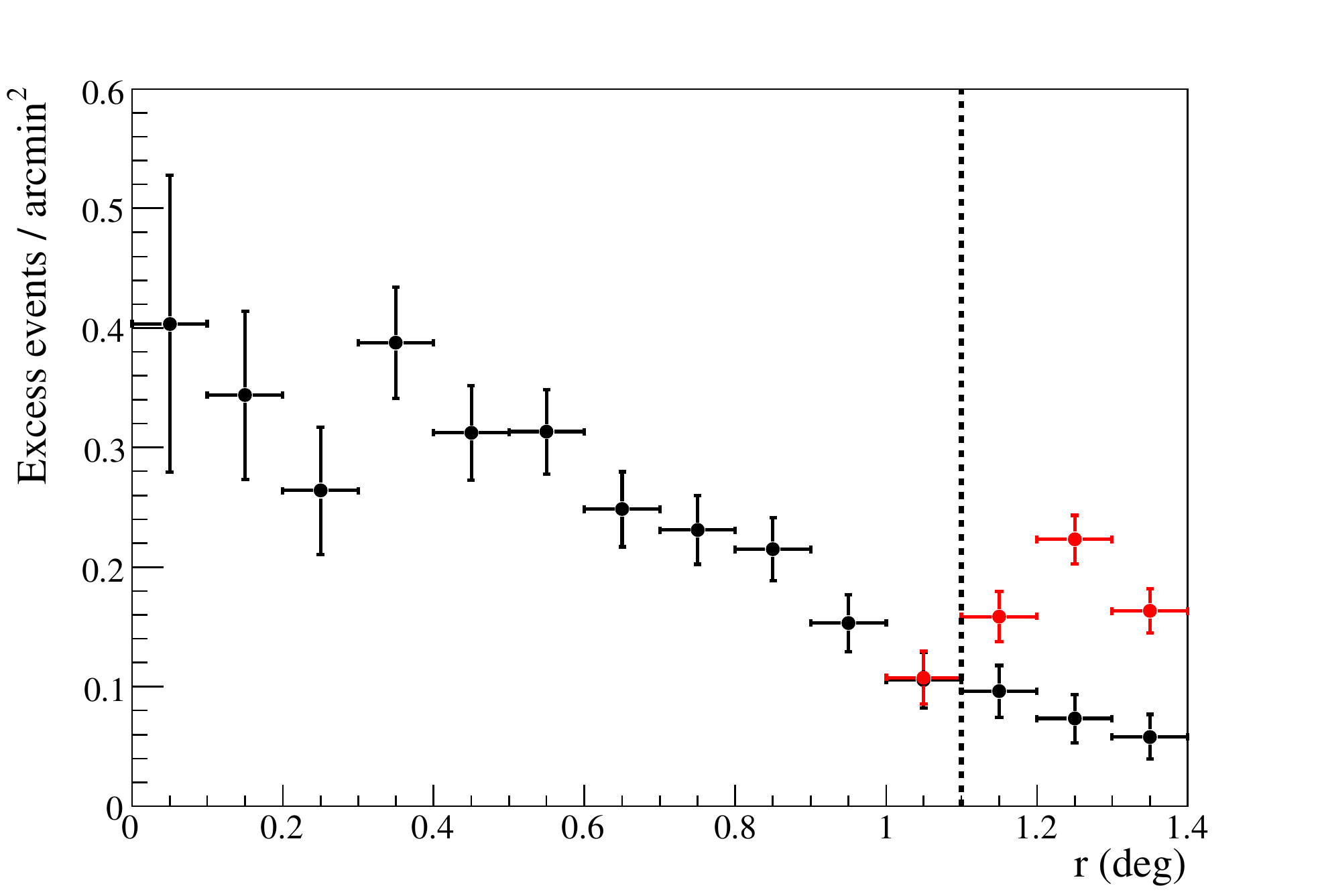}}
\caption{
H.E.S.S. radial profile relative to the best-fit position of the VHE $\gamma$-ray 
emission. The dotted vertical line denotes the 95\% containment radius used to obtain 
the spectrum shown in Figure~\ref{fig:spectrum}. Note that the region covering the 
bright source HESS~J1640--465 (see Figure~\ref{fig:VHE}) has been excluded for the 
radial profile shown in black by omitting a circle segment with 
$220^\circ \leq \phi \leq 260^\circ$ for radii of $1.0^\circ \leq r \leq 1.4^\circ$. 
The red graph displays the radial profile without excluding HESS~J1640--465.}
\label{fig:profiles}
\end{figure}

Figure~\ref{fig:VHE} shows a background-subtracted, camera
acceptance-corrected image of the VHE \g-ray counts per arcmin$^2$ of
the 3\degree $\times$ 3\degree\ FoV centred on the best fit position
of the \g-ray excess as obtained with the \emph{template} background
method. The acceptance correction has been performed using \g-ray like
background events that pass the \g-ray selection cuts. The map is
smoothed with a Gaussian kernel with a variance of 0.12\degree\ to
reduce the effect of statistical fluctuations and to highlight
significant morphological features. Significance contours from
4$\sigma$ to 8$\sigma$ are overlaid after integrating events within a
radius of 0.22\degree\ at each trial source position. This integration
radius is matched to the RMS of the Gaussian to resample significant
features in the sky image and is chosen \emph{a priori} to match the
integration radius typically used in the GPS analysis for the search
of slightly extended sources \citep{HESS:SP}. Given the extended and
complex morphology of the VHE \g-ray emission the position obtained
from a two-dimensional Gaussian fit convolved with the \hess\ PSF to
the raw excess count map obtained for \emph{\zetahard-cuts} is used to
derive an estimate on the centre of gravity of the emission. The
two-dimensional Gaussian fit gives a best fit position of RA
\HMS{16}{46}{50}$\pm 27^{\mathrm{s}}$ and Dec \DMS{-45}{49}{12}$\pm
7'$ (J2000).  Within statistical errors the centre of gravity of the
VHE \g-ray emission is consistent with the nominal Wd~1 cluster
position of RA \HMS{16}{47}{00.40} and Dec \DMS{-45}{51}{04.9}
(J2000). Based on the radial profile shown in
Figure~\ref{fig:profiles} the 95\% containment radius of the VHE
\g-ray emission relative to the best fit position is determined to be
1.1\degree. This radius is used to extract the energy spectrum
presented in Section~\ref{subsection:spectrum}.  Note that although
the sky image gives the impression that the region used for spectral
reconstruction is contaminated by \g\ rays from HESS~J1640--465, this
is mostly an artifact of the smoothing procedure. The real
contribution is less than 10\% in a ring between 1.0 and 1.1 degree
from the best fit position and only 0.8\% in the whole spectral
extraction region. Within the integration region of 1.1\degree\ a
total of $2771 \pm 139$ \g-ray excess events at a significance level
of $20.9\,\sigma$ pre-trials ($20.1\,\sigma$ post-trials) are found.

\subsection{Morphology}\label{subsection:morphology}

\begin{figure*}[t]
  \centering
  \resizebox{\hsize}{!}{\includegraphics{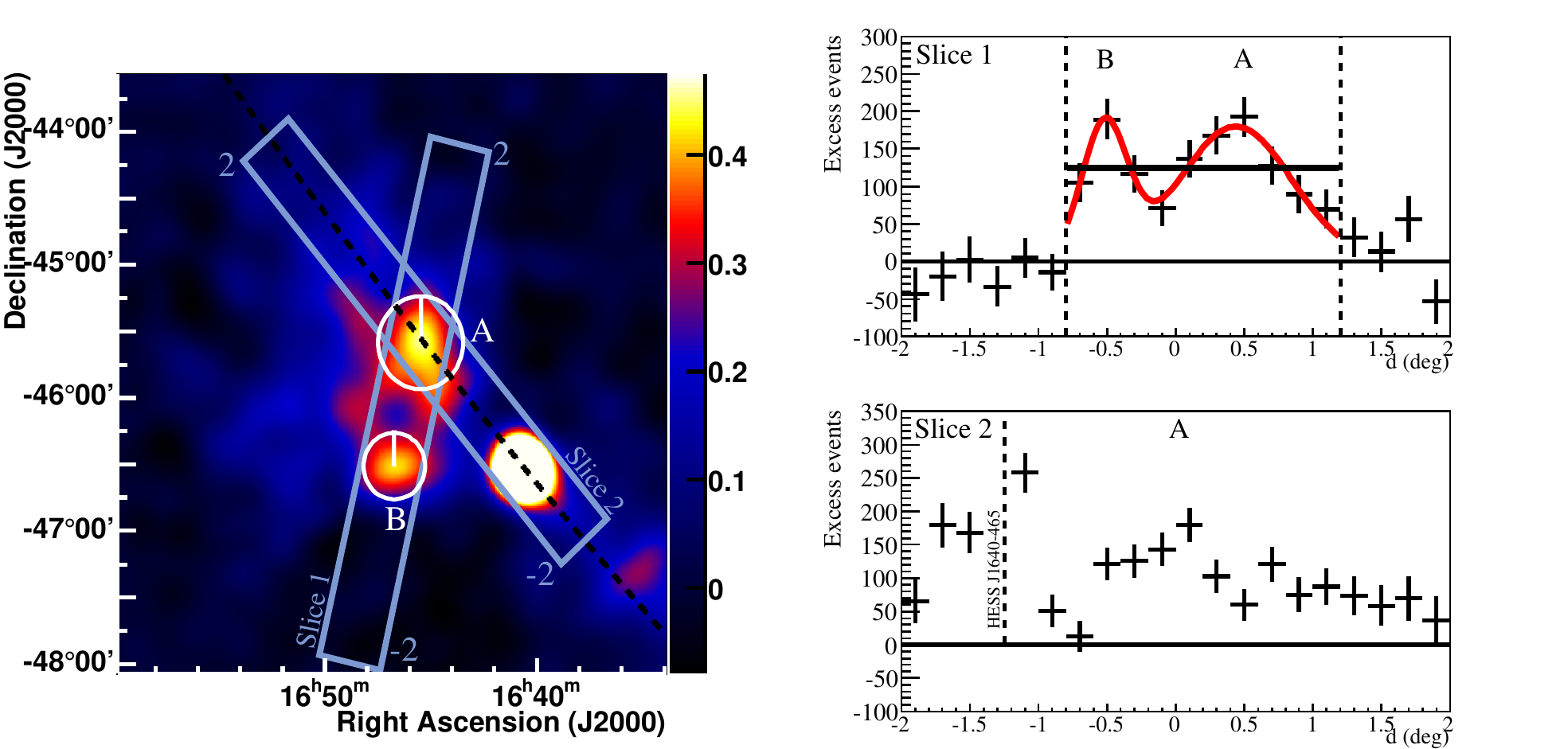}}
  \caption{\emph{Left:} \hess\ excess map as shown in Figure~\ref{fig:VHE} but 
for an enlarged region of 4.5\degree$\times$4.5\degree. The blue-grey boxes denote 
the regions used to generate the one-dimensional slices shown on the right. The 
white circles denote regions \emph{A} and \emph{B} which were used for spectral 
reconstruction (Table~\ref{tab:intflux}). The weak \g-ray emission seen in the lower 
right corner next to HESS~J1640--465 is HESS~J1634--472, also detected during the GPS 
\citep{HESS:SP}. \emph{Right, top:} Distribution of VHE $\gamma$-ray excess events in 
the blue-grey box, oriented along the two emission regions associated to \HESSJ\ and 
starting at low Declination angles. The results of a fit of a constant and of two 
sources with Gaussian shape are indicated as black and red line, respectively. 
\emph{Right, bottom:} Same as top, but oriented along the Galactic plane, starting at 
low Right Ascension angles, close to the bright source HESS~J1640--465 at 
$d=-1.25^\circ$. The slice has been truncated at 350 excess events in order to 
emphasize the VHE \g-ray emission from \HESSJ.}
  \label{fig:slices}
\end{figure*}

In order to investigate the multi-source hypothesis two emission
regions \emph{A} and \emph{B} (shown in Figure~\ref{fig:slices}
(left)) are considered. The radii of 0.35\degree\ and 0.25\degree\ of
region $A$ and $B$, respectively, are chosen according to the widths
of the two substructures. A one-dimensional slice in the uncorrelated
excess image along the major axis between the two regions has been
produced. The fit of two separate sources with Gaussian shape results
in a $\chi^2$ of 2.0 for 4 degrees of freedom with a probability of
74\%.  The probability that the emission is explained by a single
Gaussian profile or a constant value is found to be rather low at
0.2\% and 0.1\%, respectively. An $F$-test also supports the
multi-source hypothesis, given that the probabilities that the
constant or single Gaussian emission models are preferred over the
double Gaussian fit are $<0.02$ and $<0.01$, respectively.

Figure~\ref{fig:slices} (left) also suggests a contribution from
diffuse VHE $\gamma$-ray emission along the Galactic plane which
extends 1\degree\ to 2\degree\ from region \emph{A}
north-eastwards. This impression is supported by the one-dimensional
slice shown in Figure~\ref{fig:slices} (bottom right), where the
significance of the emission in all bins with distance $0.5^\circ\leq
d \leq1.8^\circ$ from the centre of region $A$ is between $2\,\sigma$
and $4\,\sigma$. This diffuse emission could be due to unresolved VHE
\g-ray sources or a Galactic diffuse emission component, caused by the
interaction of GCRs with molecular material located along the Galactic
plane. As will be shown later in Figure~\ref{fig:HI_CO-map}, there is
indeed molecular material located in this region which could act as
target for the interaction with CRs \citep[as described in
e.g.][]{Casanova09} and could account for part of the observed
emission. The statistics of region \emph{A} and \emph{B} compared to
the entire source region as obtained with the \emph{template}
background method are given in Table~\ref{tab:stats}.

\begin{table*}
  \caption{\label{tab:stats} VHE \g-ray statistics for the regions shown in 
    Figure~\ref{fig:VHE} and \ref{fig:slices}.}
  \centering
  \begin{tabular}{l c c c c c c c c}
      \hline\hline
      Region 
      & RA (J2000)
      & Dec (J2000)
      & $\theta$ 
      & \emph{On} 
      & \emph{Off} 
      & $\alpha$ 
      & Excess 
      & Significance
      \\
      
      {} 
      & deg 
      & deg
      & deg
      & events 
      & events 
      &  
      & events 
      & $\sigma$  \\
      \hline
      Full & 251.856 & -45.909 & 1.1 & 19032 & 1107471 & 0.014682 & 2771 & 20.1  \\
      $A$ & 251.370 & -45.585 & 0.35 & 2313 & 120104 & 0.014998 & 511 & 10.0 \\
      $B$ & 251.682 & -46.513 & 0.25 & 1149 & 58995 & 0.014876 & 271 & 6.5 \\
      \hline\hline
    \end{tabular}
\end{table*}

The studies presented here show some evidence for a multi-source
morphology and a separation into multiple VHE \g-ray sources.
Moreover, spectral variations across the whole emission region,
e.g. as observed for HESS~J1825--137 \citep{HESS:1825_Edep} or in the
case of the \hess\ sources in the Westerlund 2 field
\citep{HESS:Wd2_2}, could be apparent, which would further support the
multi-source hypothesis. In this case, an energy-dependent morphology
can be expected. To test this hypothesis, the complete data set has
been divided into a low-energy band, containing events with
reconstructed energies $E<1.0$\,TeV and a high-energy band, containing
events with reconstructed energies $E>1.0$\,TeV. The unsmoothed excess
maps in coarse bins of 0.3\degree\ width are used to test the
underlying distribution. A $\chi^2$ test is performed using the number
of excess events in each bin and reveals a value of 95.4 for 76
degrees of freedom (four bins covering HESS~J1640--465 have been
excluded in the calculation). Prior to the test a $\chi^2$ probability
$p_0$ of 0.05 is defined to accept/reject the null hypothesis. The
$p$-value of the test is $6.5\% > p_0$, such that the null hypothesis
that both excess maps follow the same underlying distribution cannot
be rejected. Although no energy-dependent morphology can be detected
from this test, a multi-source origin is preferred given the low
probabilities for the single source fits.

\subsection{Spectrum} \label{subsection:spectrum}

\begin{figure}[t]
  \resizebox{\hsize}{!}{\includegraphics{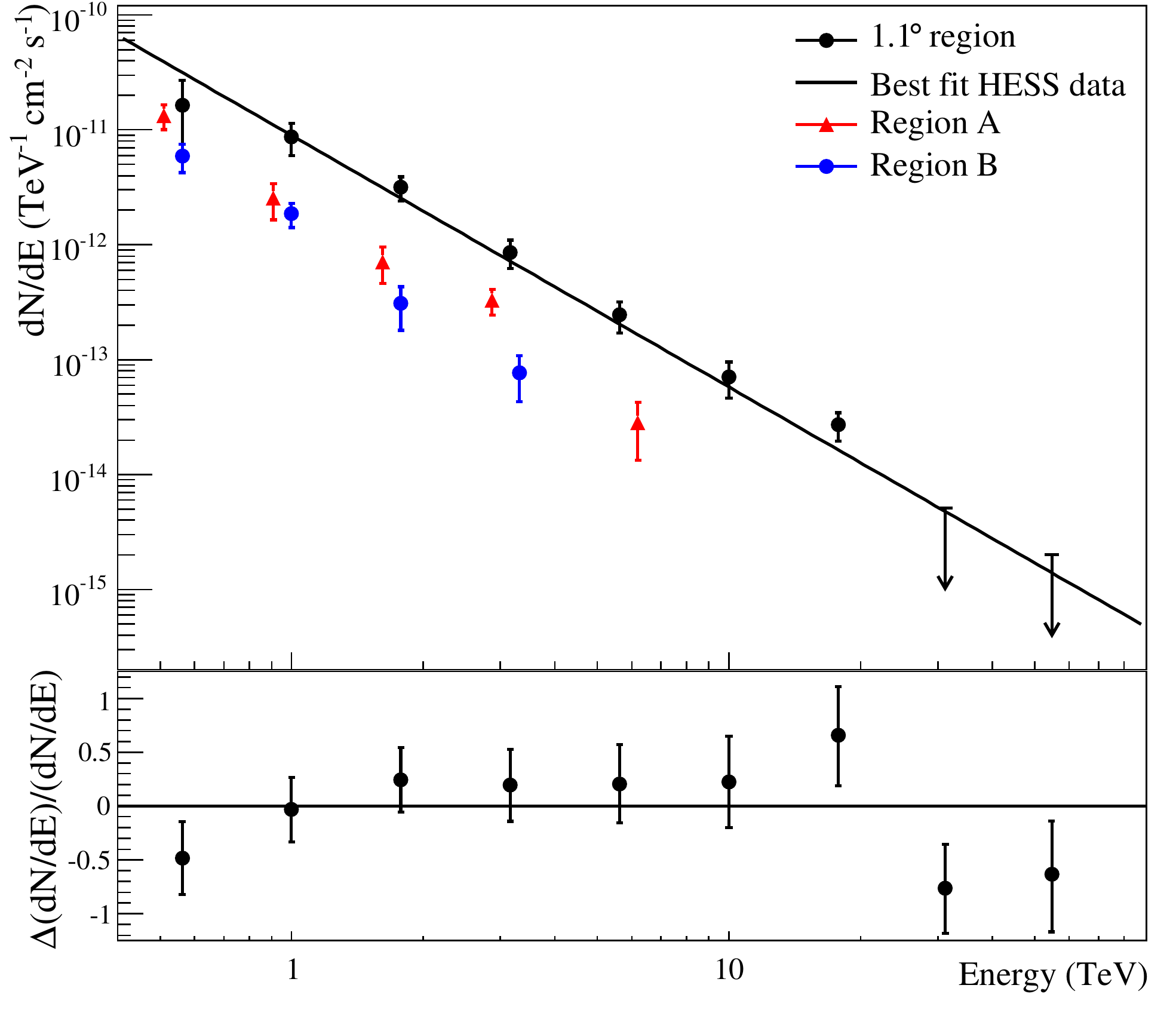}}
  \caption{\emph{Top:} Differential VHE $\gamma$-ray energy spectrum of \HESSJ. The 
data are fit by a power law $dN/dE=\Phi_0\times(E/1~\mathrm{TeV})^{-\Gamma}$. Arrows 
indicate the 95\% upper limits for spectral bins which are compatible with a zero 
flux within $1\sigma$. Also shown are spectra as obtained for region $A$ and $B$. 
\emph{Bottom:} Residuals of the power-law fit for the 1.1\degree\ \g-ray emission 
region.}
\label{fig:spectrum}
\end{figure}

The spectrum obtained for the whole emission region using the
\emph{On-Off} background estimation method is shown in
Figure~\ref{fig:spectrum}. In the fit range between 0.45\,TeV and
75\,TeV, the spectrum is well described by a power law:
$dN/dE=\Phi_0\times(E/1~\mathrm{TeV})^{-\Gamma}$ with a photon index
$\Gamma=2.19\pm0.08_{\mathrm{stat}}\pm0.20_{\mathrm{sys}}$, and a
differential flux normalisation at 1\,TeV of $\Phi_0$ =
$(9.0\pm1.4_{\mathrm{stat}}\pm1.8_{\mathrm{sys}})\times~10^{-12}\UNITS{TeV^{-1}\,cm^{-2}\,s^{-1}}$.
This translates into an integral flux above 0.2\,TeV of
$F(>0.2\,\UNITS{TeV})=(5.2\pm0.9)~\times~10^{-11}\UNITS{cm^{-2}\,s^{-1}}$. The
$\chi^2$ for the power law fit is 9.9 for 7 degrees of freedom,
yielding a $\chi^2$ probability of 19\%.

Additionally, the differential energy spectra for region \emph{A} and
\emph{B} are determined using the \emph{reflected} background
estimation method with results found to be consistent with the
\emph{On-Off} background technique. The integral flux above 0.2\,TeV
as well as the spectral results obtained from a power-law fit are
summarised in Table~\ref{tab:intflux} and compared to the results for
the spectral analysis of the whole 1.1\degree\ region. The
differential energy spectra for these two regions are shown in
Figure~\ref{fig:spectrum} as well. Within statistical errors, no
change in photon index between the three studied regions is apparent,
further supporting the lack of energy-dependent morphology across the
source based on the current data.

\begin{table*}
  \caption{\label{tab:intflux} Spectral properties of the different TeV 
    extraction regions.}
  \centering
  \begin{tabular}{l r@{$\:\pm\:$}l r@{$\:\pm\:$}l r@{$\:\pm\:$}l r@{$\:\pm\:$}l}
    \hline\hline
    Region 
    &\multicolumn{2}{c}{$\Phi_0$(1\,TeV)} 
    &\multicolumn{2}{c}{$\Gamma$}
    &\multicolumn{2}{c}{$F(>0.2$\,TeV)} 
    &\multicolumn{2}{c}{\% total} 
    \\
    
    {} 
    & \multicolumn{2}{c}{$10^{-12}\UNITS{TeV^{-1}\,cm^{-2}\,s^{-1}}$} 
    & \multicolumn{2}{c}{}  
    & \multicolumn{2}{c}{$10^{-11}\UNITS{cm^{-2}\,s^{-1}}$} 
    & \multicolumn{2}{c}{}  \\
    \hline
    Full & 9.0 & 1.4 & 2.19 & 0.08 & 5.2 & 0.9 & \multicolumn{2}{c}{100} \\
    \emph{A} & 2.1 & 0.3 & 2.11 & 0.12 & 1.1 & 0.2 & 21 & 4 \\
    \emph{B} & 1.4 & 0.2 & 2.29 & 0.17 & 0.8 & 0.2 & 15 & 4 \\
    \hline\hline
  \end{tabular}
\end{table*}

%
%
\section{Discussion}
In this section, the spectral and morphological results are used to
elaborate possible acceleration scenarios related to \HESSJ. Although
the morphological analysis prefers a two-source approach (at the
$\sim2.5\,\sigma$ level), the similarity of the spectra from region
\emph{A} \& \emph{B} does not allow to substantiate the preference for
a two-source scenario further. Owing to this ambiguity, we pursue both
possibilities through all portrayed counterpart scenarios.

\begin{figure}[t]
  \resizebox{\hsize}{!}{
    \includegraphics{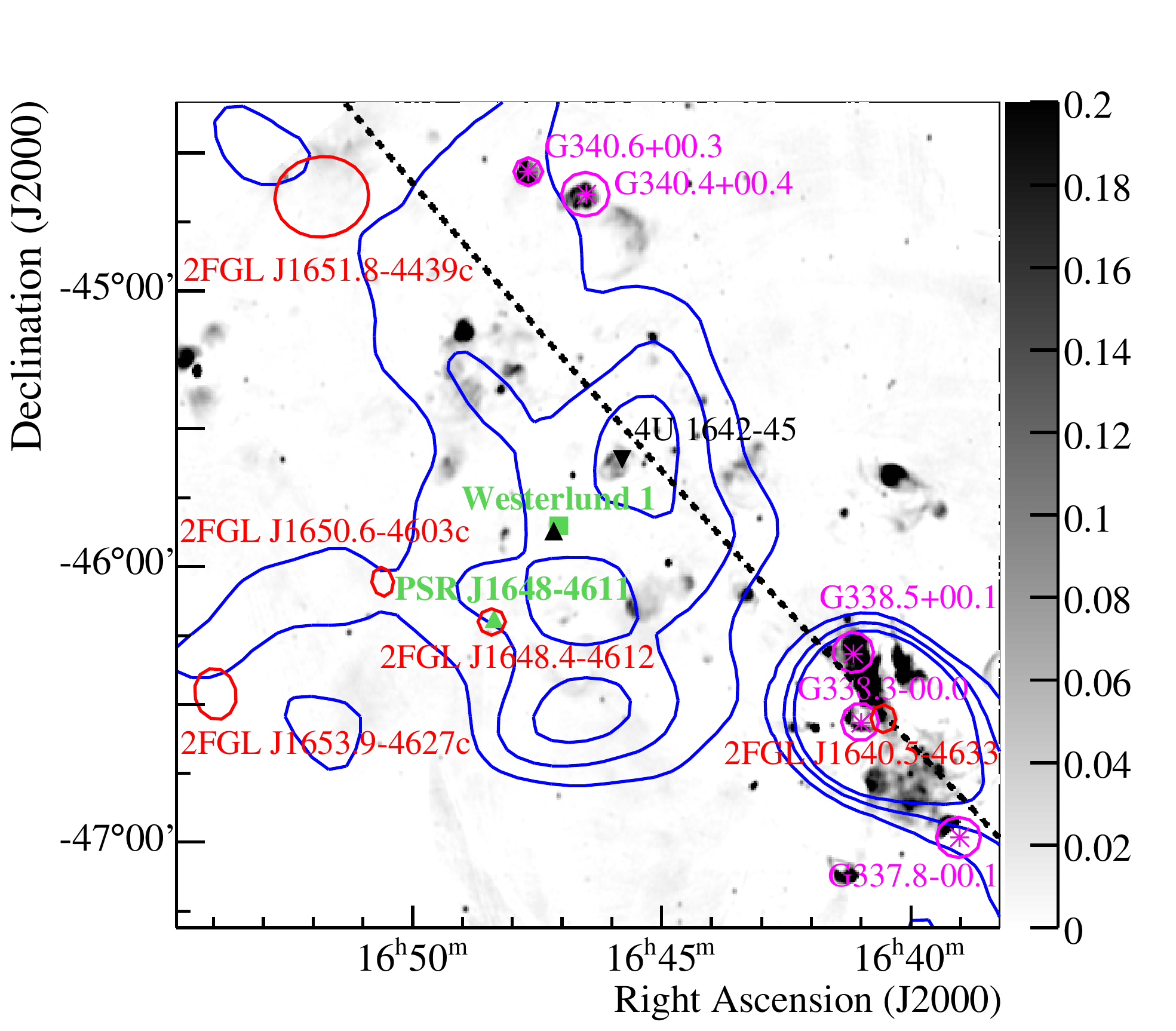}}
  \caption{\hess\ smoothed VHE \g-ray excess contours in blue overlaid
    on the Molonglo 843~MHz map \citep[grey scale, in
    Jy/beam;][]{Molonglo_Cat}. Also shown are SNRs
    \citep[purple,][]{Green_SNRs}, the pulsar PSR~J1648--4611
    \citep{ATNF_Cat}, the LMXB 4U~1642--45 \citep{Forman1978+},
    \emph{Fermi}-LAT sources \citep{Fermi:2FGL}, the magnetar
    CXOU~J164710.2--455216 \citep{Muno06a} as a black upright triangle
    and the stellar cluster Wd\,1.}
  \label{fig:vicinity}
\end{figure}

Due to the large size of \HESSJ\ ($\sim2^\circ$), a few objects that
could be regarded as potential VHE \g-ray emitter are found in the
region of interest. Besides the massive stellar cluster Westerlund 1,
a high spin-down power pulsar PSR~J1648--4611 \citep{ATNF_Cat} which
has recently been discovered to be a high-energy $\gamma$-ray pulsar
with a possible unpulsed $\gamma$-ray component is found
\citep{Kerr2009,Fermi1year2010,Fermi:2FGL}. Furthermore, the low-mass
X-ray binary (LMXB) 4U~1642--45 \citep{Forman1978+}, the magnetar
candidate CXOU~J164710.2--455216 \citep{Muno06a} and three
unidentified \emph{Fermi}--LAT sources, 2FGL~J1650.6--4603c,
2FGL~J1651.8--4439c and 2FGL~J1653.9--4627c \citep{Fermi:2FGL} are
located within \HESSJ\ (Fig. \ref{fig:vicinity}).

In the following, the discussion will focus only on the
astrophysically associated objects with known distance and energetics.

\subsection{4U~1642--45} 
The X-ray bright LMXB 4U~1642--45 with an accreting neutron star lies
within the strong emission region \emph{A} (Fig. \ref{fig:vicinity})
and is located at a distance of 8.5 to 11.8~kpc
\citep{vanParadijs1995+,Grimm2002+,Gilfanov2003+}. It exhibits an
average X-ray luminosity of $L_\mathrm{X} \approx
10^{38}$\,erg\,s$^{-1}$ and \citet{Grimm2002+} reported of periods
of super-Eddington luminosity from \emph{ASM} observations.

So far, only high-mass X-ray binary systems have been associated with
point-like and variable VHE $\gamma$-ray sources
\citep[e.g. LS~5039,][]{HMXB_Aharonian2005+,Aharonian2006+LS5039}. At
8.5 to 11.8~kpc, \HESSJ\ would be 320 to 450~pc in size and region
\emph{A} 100 to 144~pc, for angular diameters of 2.2$^\circ$ and
0.7$^\circ$ respectively. These inferred source sizes are well beyond
the \hess\ PSF in this analysis (26 to 35~pc in diameter). A check for
variability at the position of 4U~1642--45 did not reveal any
indication of such as has been observed, e.g. in the case of LS~5039
and HESS~J0632+057
\citep{Aharonian2006+LS5039,Aharonian2007+Monoceros}.

In summary, an association of 4U~1642--45 and the entire VHE
$\gamma$-ray emission region is unlikely. The extent of the subregion
\emph{A} also disfavours a scenario with 4U~1642--45 and a subregion
as well. An association with region \emph{A} would be solely based on
spatial coincidence but would then present a new class of VHE
$\gamma$-ray sources.

\subsection{CXOU~J164710.2--455216}
This anomalous X-ray pulsar is considered a magnetar and has been
associated with Wd~1 given its apparent proximity to the cluster and
the low probability of a random association \citep{Muno06a}. Shortly
after its X-ray discovery, an outburst in X-rays was reported from
\emph{Swift} observations \citep{Campana2006+}. Additionally,
XMM-\emph{Newton} observations were conducted prior to and after the
outburst \citep{Muno2007+,Israel2007+}.

The observed X-ray luminosity is
$L_\mathrm{X}\approx3\times10^{33}$\,erg\,s$^{-1}$ which increased
during the outburst by a factor of $\sim100$. The rotation period is
10.6\,s with a derivative of $\dot{P}=9.2\times10^{-13}$\,s\,s$^{-1}$
which infers an X-ray spin-down power of
$\dot{E}=3\times10^{31}\,$erg\,s$^{-1}$ and a characteristic age of
$\tau=1.8\times10^5$\,yrs. The surface magnetic-field strength is
$\sim1\times10^{14}$\,G. The relevant properties of
CXOU~J164710.2--455216 are listed in Table \ref{tab:pulsars}.

\begin{table*}
  \caption{\label{tab:pulsars} Properties and inferred \g-ray
    luminosities of the pulsars and the VHE \g-ray emission regions.}
  \centering
    \begin{tabular}{l c c c c c c}
      \hline
      \hline
      Object / Region & $\tau$ & $d$ & $P_0$ & $\dot{P}$& $\dot{E}$ & $L_\gamma$\tablefootmark{(1)} \\
      & $10^5$\,yrs & kpc & s & $10^{-13}$\,s\,s$^{-1}$ & $10^{34}$\,erg\,s$^{-1}$ & $10^{34}$\,erg\,s$^{-1}$\\
      \hline
      CXOU~J164710.2--455216\tablefootmark{a}&1.8&5&10.6&9.2&$3\times10^{-3}$&26\\
      PSR~J1648--4611\tablefootmark{b}&1.1&5.7&0.2&0.2&21&34\\
      \hline
      \HESSJ\ &--&4.3&--&--&--&19\\
      Region \emph{A}&--&4.3&--&--&--&4.6\\
      Region \emph{B}&--&4.3&--&--&--&2.9\\
      \hline
    \end{tabular}
    \tablefoot{
      \tablefoottext{1}{Obtained by scaling the observed VHE \g-ray flux between
        0.1 and 100\,TeV to the nominal or adopted distance of the respective
        object.}
    }
    \tablebib{
      \tablefoottext{a}{\citet{Muno06a,Israel2007+}.}
      \tablefoottext{b}{\citet{ATNF_Cat}.}
    }
\end{table*}

Since the X-ray luminosity exceeds the rotational spin-down power,
this suggests that the observed X-ray emission is not due to spin-down
processes, but some other form. For magnetars, the intense magnetic
field can yield considerable power in magnetic dissipative
processes. Following \citet{Zhang2003}, such power can be of the order
$L_\mathrm{mag} \approx 10^{35}$\,erg\,s$^{-1}$, which, for
CXOU~J164710.2--455216, would dominate over the rotational spin-down
power. If acceleration of particles to TeV energies occurs due to
magnetic energy release one could expect a rather compact TeV-emission
region similar to spin-down powered TeV pulsar wind nebulae (PWNe)
like the Crab nebula \citep{HESS:Crab}.

However, it is still under investigation if magnetars can exhibit PWNe
in general. For most pulsars, the respective PWNe would be too faint
to be detected in X-rays \citep{Gaensler2006+}. In the VHE \g-ray
regime, young ($\mathcal{O}(10^3$\,yrs$)$) and high spin-down power
($\mathcal{O}(10^{34}$\,erg\,s$^{-1})$) magnetars seem to be promising
objects \citep{Halpern2010+}. 

However, the rotational spin-down power of CXOU~J164710.2--455216 is
too low to account for both the entire observed VHE $\gamma$-ray
emission and either subregion. Therefore, any acceleration of
particles would have to involve a magnetic energy release with a power
output of $L_\mathrm{mag}\ge10^{35}$\,erg\,s$^{-1}$. The observed
emission could then be explained by an energy conversion process that
operates with efficiency up to 100\% but the size of \HESSJ\
($\sim$160\,pc at the distance of the magnetar) stands in
contradiction to the expected compact region. Either of the two
subregions could be energetically explained with efficiencies in the
order of 0.5 but again as one would expect the resulting PWN to be
compact and close to the magnetar itself, these scenarios are
disfavoured as well.

All in all, VHE $\gamma$ rays from CXOU~J164710.2--455216 is not a
favoured scenario to account for \HESSJ\ or parts from it as its
relevant properties and current (V)HE $\gamma$-ray observations do not
support such an approach.

\subsection{PSR~J1648--4611}
Among known VHE $\gamma$-ray sources, PWNe are the most abundant
source type: Roughly one third of these sources are associated with
PWN systems \citep[e.g.][]{Hinton2009+}. The pulsar is found to be
located in the centre of the nebula \citep[e.g. the Crab
nebula,][]{HESS:Crab} or offset from it \citep[e.g. the Vela X
nebula,][]{HESS:VelaX}. Most of the VHE \g-ray-emitting PWNe are
spatially extended and offset from the pulsar position and efficient
in terms of converting available spin-down power into $\gamma$-ray
emission. In general, PWNe firmly associated with known pulsars
convert 10\% at most of their spin-down power into $\gamma$-ray
luminosity \citep{Gallant2007}.

The high spin-down power pulsar PSR~J1648--4611 with
$\dot{E}=2.1\times10^{35}$\,erg\,s$^{-1}$ \citep{ATNF_Cat} is located
within the emission region \HESSJ\ and associated with the
\emph{Fermi}-LAT source 2FGL~J1648.4--4612 \citep{Fermi:2FGL}. Pulsed
$\gamma$-ray emission with the rotation period of PSR~J1648--4611 as
well as signatures for a constant (possible PWN) contribution have
been reported. Therefore, this object has to be considered a
$\gamma$-ray pulsar at low GeV energies. Inspection of archival
\emph{Chandra} data with 10\,ks exposure (obsid: 11836) shows no
indication for the presence of an X-ray counterpart to
PSR~J1648--4611. Following the relation between spin-down power and
X-ray luminosity of PWNe \citep{Mattana2009+}, the expected X-ray
luminosity $L_\mathrm{X}\approx5\times10^{30}$\,erg\,s$^{-1}$ of the
associated PWN would not be detectable with current X-ray instruments
and is hence consistent with the non-detection with \emph{Chandra}.

The first attempt to explain \HESSJ\ is a single-source scenario which
is motivated by the presence of the HE pulsar PSR~J1648--4611. The
inferred $\gamma$-ray luminosity (0.1 - 100\,TeV) at the distance of
PSR~J1648--4611 would require a conversion efficiency
$\epsilon_\mathrm{eff}\approx1$. Additionally, the size of the VHE
\g-ray-emitting region would extend over roughly 200\,pc which is a
factor 3 -- 10 larger than known extended VHE \g-ray-emitting PWN
systems. Given the size, cooling losses would lead to considerable
softening of the VHE \g-ray spectrum and a downward shift of the
maximum energy with increasing distance from the pulsar -- similar to
the spectral softening of HESS~J1825--137 \citep{HESS:J1825}. The VHE
\g-ray photon index reconstructed at the position of PSR~J1648--4611
is $\Gamma=2.37\pm0.43$ using the \emph{reflected}-background regions
method and assuming a point-like source origin. This is compatible
within statistical errors with the emission of the entire region.
However, the available statistics do not permit firm conclusions about
potential spectral changes across \HESSJ\ to be drawn. In summary, the
unprecedented high efficiency needed and the size of the VHE
$\gamma$-ray emission region disfavour \HESSJ\ as a very extended PWN.

In a two-source approach as motivated in Section
\ref{subsection:morphology}, a displacement of PSR~J1648--4611 from
either subregion is apparent. The pulsar is displaced by
$\sim50-70$\,pc from region \emph{A} and \emph{B}.  These offsets are
large compared to those of known VHE \g-ray PWNe but could in
principle be explained by relative proper motion of the pulsar
assuming a transversal velocity of \order($500$\,km\,s$^{-1}$) which
is at the upper end of the range of known transversal motions of
pulsars \citep{ATNF_Cat}. In this scenario, one of the two subregions
could in principle be powered by the pulsar with
$\epsilon_\mathrm{eff}\approx0.1\ldots 0.2$. In this case, the
morphology would reflect the ambient conditions, e.g. one of the
subregions could be the result of the expansion of the PWN into a
low-density medium or could be due to an asymmetric reverse shock of
the SNR.

Recently, \citet{Luna2010+} proposed that a region of low density in
CO seems to partially match structures in preliminary VHE \g-ray data
\citep{HESS:Wd1} and for which a SN event with PSR~J1648--4611 as
precursor could be responsible. However, the kinematic age of the
cavity is about 55 times larger than the characteristic age of
PSR~J1648--4611 and the inferred subsonic expansion velocity is
insufficient to accelerate particles up to the VHE \g-ray
regime. Moreover, other SNR candidates at the position of this cavity
are not to be found in archival data and the morphology of \HESSJ\
does not strongly motivate a shell-like structure centred at about
\HMS{16}{47}{23.3}, \DMS{-45}{42}{5.2} with an inner radius of
$\sim$0.5$^\circ$ \citep{Luna2010+}. Although it cannot be ruled out
that this cavity could be blown as a consequence of a SN it seems
unplausible that the whole $\sim25$\,pc-thick shell would expand
uniformly at 6 - 8\,km\,s$^{-1}$ and thereby giving rise to particle
acceleration up to TeV energies.

Similarly as concluded for the magnetar, \HESSJ\ seems unlikely to be
explained as a very extended PWN powered by PSR~J1648--4611. Either
one of the subregions could be explained by an offset PWN. The
inferred offsets would be comparatively large but the required
efficiency would be amongst known TeV PWNe.

\begin{figure}[t]
  \resizebox{\hsize}{!}{
    \includegraphics{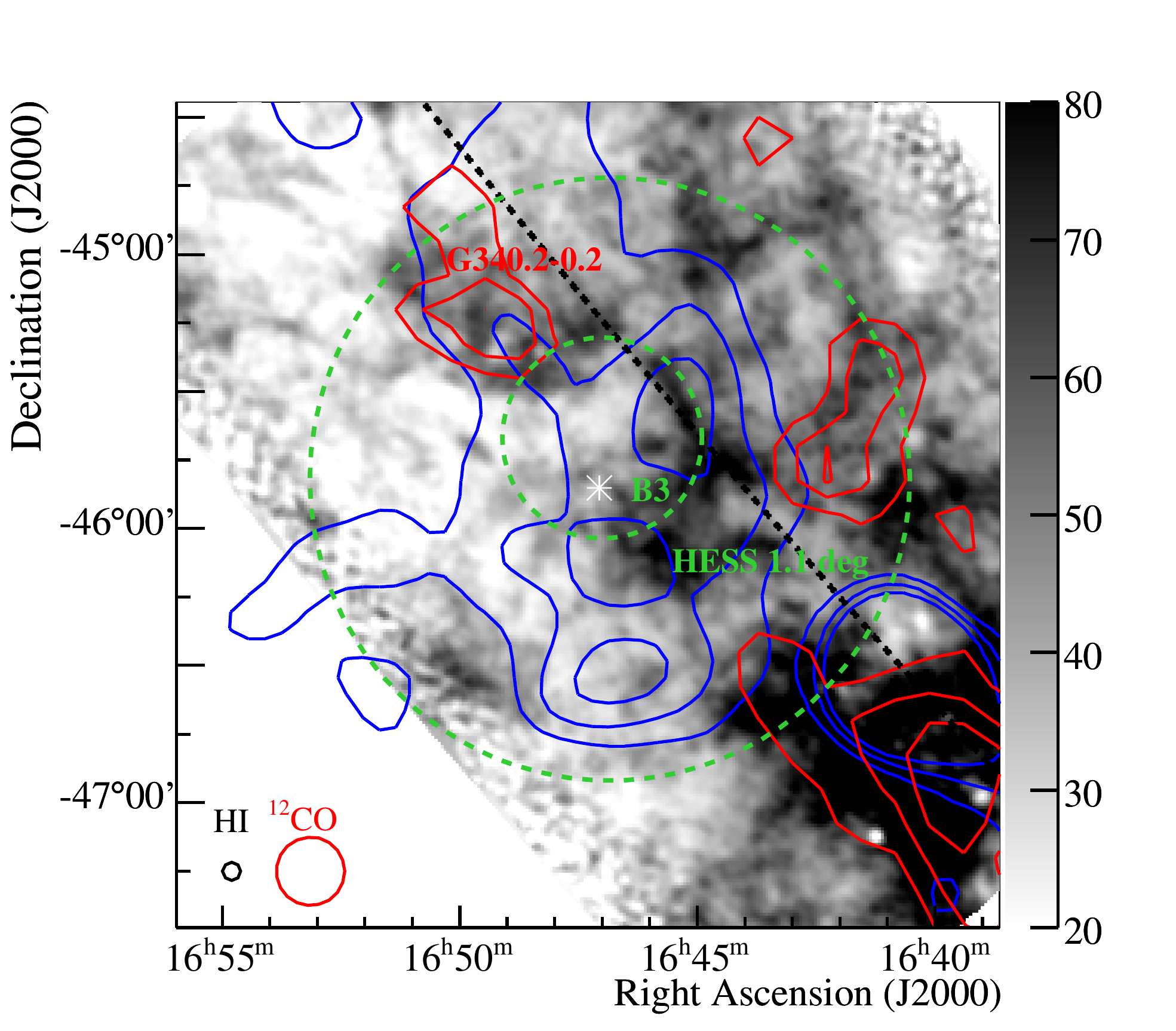}}
  \caption{HI 21\,cm line emission map at a radial velocity of about
    $-55$\,km\,s$^{-1}$ \citep{SGPS_Cat} between 20 and 80\,K. Bright
    regions are HI-depleted whereas dark regions are
    HI-dense. Overlaid are the CO contours (red) at 12.5, 22.5 and
    32.5\,K \citep{Dame2001+} along with the smoothed \HESSJ\ contours
    in blue. The estimated position of the HI void \textit{B3} is
    marked by the small green dashed circle \citep{Kothes07}. The
    large green dashed circle indicates the region used for the
    spectral reconstruction in the TeV regime and to compute the
    ambient matter density. The position of Wd~1 is marked by a white
    star in the centre. The circles in the lower left corner indicate
    the beam sizes of the respective radio observations.}
  \label{fig:HI_CO-map}
\end{figure}

\subsection{Westerlund~1}\label{subsection:wd1}
The motivation for \hess\ observations of the Wd~1 region has been
outlined earlier (see Section \ref{section:introduction}) and the
analysis of \HESSJ\ revealed that Wd~1 is located close to the
centroid of the VHE $\gamma$-ray emission. Hence this massive cluster
is an interesting object to consider. As some of its properties are
still subject of on-going investigation certain assumptions are made:
\emph{(1)} Due to the high visual extinction of $A_v \approx
12^{\mathrm{mag}}$, distance estimates based on photometry and
spectroscopy vary strongly between 1.1\,kpc and 5.5\,kpc
\citep{Westerlund87,Piatti1998+,Clark05,Brandner2008+}. \citet{Kothes07}
derived a distance of 3.9\,kpc based on the HI line emission which
\citet{Luna2010+} extrapolated to $4.3$\,kpc using the IAU distance
recommendation to the Galactic Centre of 8.5\,kpc. In this work, we
adopt the 4.3\,kpc albeit using a more recent Galactocentric distance
estimate of 8.33\,kpc \citep{Gillessen2009+}. \emph{(2)} Different
model-dependent approaches to estimate the age suffer from the
apparent lack of main-sequence stars. Recently,
\citet{Negueruela2010+} estimated the age to be $\gtrsim5$\,Myrs and
an age of 5\,Myrs is adopted here. \emph{(3)} A cluster mass of
$10^5$\MS\ is assumed.

Estimates based on the stellar population of Wd~1 imply that 80 to 150
stars with initial masses exceeding 50\,M$_{\odot}$ have already
undergone a SN explosion \citep{Muno06b}. However, neither radio nor
X-ray observations have revealed any SNR candidate. In order to
estimate the total kinetic energy dissipated by the system through SNe
and stellar winds the \emph{Starburst99} cluster evolution model
\citep{Leitherer1999+,Vazquez2005+,Leitherer2010+} has been used. The
default parameters with the standard Kroupa initial mass function
(with the exponents 1.3 and 2.3, \citep{Kroupa2002}) and the
\textit{evolutionary}-model option, have been chosen. The total energy
dissipated at the nominal age of Wd~1 including stellar winds and SNe
amounts to $E_\mathrm{kin}=3.0\times10^{53}\,\mathrm{erg}$.

For an adiabatically expanding wind \citep{Weaver77,Silich2005+}, a
structured and expanding ($\sim30$\,km\,s$^{-1}$) hot bubble with an
outer shock radius of $\sim250$\,pc is expected to form around Wd~1
even if only a moderate number of 50~OB stars are considered. However,
a dedicated search for such a super bubble-like structure in HI and CO
data at the radial velocity of Wd~1 did not reveal any signatures of
such a feature \citep{SGPS_Cat,Kothes07}. The latter authors find
indications for a much smaller (55\,pc) and slowly expanding
(3\,km\,s$^{-1}$) feature \citep[\emph{B3} in][]{Kothes07}, see
Fig.~\ref{fig:HI_CO-map}. However, in this first approximation,
radiative cooling is not considered although this cooling process
might affect the evolution of stellar cluster winds
\citep[e.g. discussed in][]{Wuensch08}.

\citet{Kothes07} interpret the formation of the HII region complex
G$340.2-0.2$ (depicted in Fig.~\ref{fig:HI_CO-map}) as triggered by
the stellar wind activity of Wd~1. Indeed, if Wd~1 is located at a
distance of 4.3\,kpc some correlation between VHE \g\ rays and the
location of this HII region is expected.

An average gas density for atomic hydrogen $n_\mathrm{HI}$ and
molecular gas $n_{\mathrm{H}_2}$ can be derived using available HI and
CO data in the velocity range of Wd~1 ($-50$ to $-60$\,km\,s$^{-1}$)
for the entire 1.1$^\circ$ region of interest. For this, all pixel
values in this velocity range and within $1.1^\circ$ from the cluster
position are considered as well as the SGPS beam size (130\,arcsecs)
and the oversampling factor (11.97). With the HI intensity-mass
conversion factor of
$1.823\times10^{18}\,\mathrm{cm}^{-2}/(\mathrm{K\,km}\,\mathrm{s}^{-1})$
\citep{Yamamoto2003+}, the result\footnote{The density $n$ is
  proportional to $\sum b^2 p_i R^{-1} f_\mathrm{s}^{-2}$ where $p_i$
  are the pixel values of the HI or CO map, $R$ the radius of the
  region of interest, $b$ the beam size of the respective experiment
  and $f_\mathrm{s}$ is the oversampling factor in the respective HI
  or CO maps used.} is $n_\mathrm{HI} = 0.24\,$cm$^{-3}$. In the CO
data \citep{Dame2001+}, a low-density region around the cluster seems
to be apparent as well (Fig.~\ref{fig:HI_CO-map}). This feature is
larger than \textit{B3}. With a beam size of 450\,arcsecs and a
oversampling factor of 1, the application of the conversion factor of
CO to H$_2$ of
$1.5\times10^{20}\,\mathrm{cm}^{-2}/(\mathrm{K\,km}\,\mathrm{s}^{-1})$
\citep{Strong2004+} leads to an average density of H$_2$ molecules in
units of atomic hydrogen of $n_{\mathrm{H}_2}=12.16\,$cm$^{-3}$. Note
that the X factor used for the CO data incorporates caveats pointed
out in \citet{Strong2004+}. The required energy in CRs to power the
$\gamma$-ray emission can now be estimated:
\begin{eqnarray}
  E_\mathrm{CR} & = & 2.1\times10^{50} \left(\frac{L_{>1\,\mathrm{GeV}}}{5.8\times10^{35}\,\mathrm{erg\,s}^{-1}}\right)\left(\frac{n_\mathrm{HI}+n_{\mathrm{H}_2}}{12.4\,\mathrm{cm}^{-3}}\right)^{-1}\,\mathrm{erg}\,, \nonumber
\end{eqnarray}
where $L_{>1\,\mathrm{GeV}}$ is the high-energy luminosity computed
with the spectral results presented earlier between 1\,GeV and
1\,PeV. 

When interpreted in a single-source scenario, the required efficiency
for transferring kinetic energy in shocks or turbulences into
energetic particles through acceleration is therefore at the level of
$\sim0.1$\%.  For a similar argument as for the PWN interpretation, a
leptonic origin of the VHE $\gamma$-ray emission is difficult to
reconcile with the large extent of \HESSJ\ which translates into a
size of 160\,pc at the distance of the cluster. Given the large photon
density in the environment of Wd~1, a fast convective transport of the
electrons would be required to prevent them from cooling. Hence a
dominant hadronic origin is favoured in this approach.

A possible two-source scenario would also consist of a dominant
hadronic CR component as the stellar photon field would lead to a
rapid cooling of VHE electrons. In this case, bright VHE $\gamma$-ray
structures would trace dense features in HI and CO.  In particular,
region \emph{A} lies at the edge of \emph{B3} and is coincident with
dense structures in HI (see Fig. \ref{fig:HI_CO-map}) which would
naturally provide sufficient target material. Region \emph{B},
however, remains comparatively dark as HI and CO data do not suggest a
higher abundance of target material.

In summary, Wd~1 and its massive stars favour a hadronic mechanism for
the entire emission region.  Here, the size and the inferred
energetics of \HESSJ\ could be plausibly explained. In case of the two
subregions, or multiple source regions in general, more observational
data in all wavelength bands and detailed modelling are required.

%
%
\section{Summary and Conclusions}
In summary, \HESSJ\ is a new VHE \g-ray source found towards the
unique massive stellar cluster Westerlund~1 and a number of other
potential counterparts. The large size of \HESSJ\ however, over 2
degrees in diameter making it one of the largest TeV sources so far
detected by H.E.S.S., presents a challenge in identifying a clear
counterpart (or a number of counterparts) to explain the VHE \g-ray
emission.  

The detection of degree-scale VHE $\gamma$-ray emission, namely
\HESSJ, towards the stellar cluster Westerlund 1 with a total
significance of 20\,$\sigma$ from H.E.S.S. observations performed in
the years 2004, 2007 and 2008 (33.8~hrs live time) is reported. The
energy spectrum between 0.45\,TeV and 75\,TeV of the entire region is
best fit by a simple power law with an index $\Gamma = 2.19 \pm
0.08_\mathrm{stat} \pm 0.2_\mathrm{sys}$ and a normalisation at 1\,TeV
$\Phi_0 = (9.0 \pm 1.4_\mathrm{stat} \pm 1.8_\mathrm{sys}) \times
10^{-12}\,$TeV$^{-1}$cm$^{-2}$s$^{-1}$ with $\chi^2/\mathrm{ndf} =
9.9/7$. The integrated flux above 0.2\,TeV amounts to $(3.49 \pm 0.52)
\times 10^{-11}\,$cm$^{-2}\,$s$^{-1}$. The VHE $\gamma$-ray luminosity
between 0.1 and 100\,TeV is
$1.9\times10^{35}\,(d/4.3\,\mathrm{kpc})^2\,$erg\,s$^{-1}$.

The centroid of \HESSJ\ is consistent with the nominal position of
Wd~1. The observed VHE $\gamma$-ray emission region has a diameter of
about $2^\circ$ which translates into a spatial extent of 160\,pc at
the distance of Wd~1 or 200\,pc at the distance of PSR~J1648--4611. In
either case, the size of \HESSJ\ would be the largest among currently
known VHE $\gamma$-ray sources, if the emission is of a single-source
origin. This is supported by the lack of spectral changes across
\HESSJ\ within statistical errors. Although there is some evidence for
a multi-source morphology, the limited statistics hamper a detailed
investigation into the presence of multiple sources.

In a scenario where one astrophysical object is responsible for
\HESSJ, Wd~1 could naturally account for the required injection power
provided that about 0.1\% ($n/12.4\,$cm$^{-3}$) of the kinetic energy
released by stellar winds and supernova explosions are converted into
particle acceleration. In this case, the stellar wind and SNe activity
of Wd~1 would strongly affect the surroundings but also reflect the
ambient conditions.

In a split of \HESSJ\ into two distinct subregions, however, a
superposition of two or more sources adding up to the observed VHE
$\gamma$-ray emission region could be possible. In this case, one of
the subregions could also be explained by a PWN with comparatively
large offset from PSR~J1648--4611. Despite the spatial coincidence,
the LMXB 4U~1642--45 and the magnetar CXOU~164710.2--455216 are not
likely to account for \HESSJ\ or parts of it.

Further multiwavelength observations (in particular those covering the
VHE \g-ray source in full), and deeper VHE \g-ray coverage will no
doubt be valuable in shedding light on the nature of this source.

\begin{acknowledgements}
  The support of the Namibian authorities and of the University of
  Namibia in facilitating the construction and operation of \hess\ is
  gratefully acknowledged, as is the support by the German Ministry
  for Education and Research (BMBF), the Max Planck Society, the
  French Ministry for Research, the CNRS-IN2P3 and the Astroparticle
  Interdisciplinary Programme of the CNRS, the U.K. Particle Physics
  and Astronomy Research Council (PPARC), the IPNP of the Charles
  University, the South African Department of Science and Technology
  and National Research Foundation, and by the University of
  Namibia. We appreciate the excellent work of the technical support
  staff in Berlin, Durham, Hamburg, Heidelberg, Palaiseau, Paris,
  Saclay, and in Namibia in the construction and operation of the
  equipment. This research has made use of NASA's Astrophysics Data
  System. This research has made use of the SIMBAD database, operated
  at CDS, Strasbourg, France. SO acknowledges the support of the
  Humboldt foundation by a Feodor-Lynen research fellowship.
\end{acknowledgements}

\bibliographystyle{aa.bst} 
\bibliography{wd1_paper_AA_revised} 
\end{document}